\documentclass[english,12pt]{article}
\usepackage[T1]{fontenc}
\usepackage[utf8]{inputenc}
\usepackage[english]{babel}
\usepackage[pdftex]{graphicx}
\usepackage[width=\textwidth]{caption}
\usepackage[top=1in, bottom=1in, left=1in, right=1in]{geometry}
\usepackage{amsmath,amsopn,amssymb,bm, dsfont,mathrsfs}
\usepackage{a4wide}
\usepackage{array,float, placeins,flafter, longtable,booktabs,pdflscape}
\usepackage{verbatim}
\usepackage{color}
 \usepackage[sectionbib]{bibunits}
\defaultbibliographystyle{chicago}
\defaultbibliography{biblio}
 \usepackage{natbib}
\usepackage{lmodern}
\usepackage{mathtools}
\usepackage{multirow}
\usepackage[hidelinks]{hyperref}
\usepackage{comment}
\usepackage{subfig}
\usepackage{graphicx}

\makeatletter

\newtheorem{thm}{Theorem}

\newtheorem{hyp}{Assumption}

\newtheorem{des}{Design}

\newcommand{\TE}{\text{TE}}

\newcommand{\Supp}{\text{Supp}}

\newcommand{\WAS}{\text{WAS}}

\newcommand{\R}{\mathbb R}

\newcommand{\ind}[1]{\mathds{1}\left\{#1\right\}}
\newcommand{\eqd}{\stackrel{d}{=}}

\newcommand{\Cov}{\text{Cov}}
\newcommand{\eps}{\varepsilon}

\newcommand{\convL}{\stackrel{d}{\longrightarrow}}
\newcommand{\convP}{\stackrel{p}{\longrightarrow}}

\newcommand{\st}[1]{\texttt{#1}}

\renewcommand{\section}{\@startsection{section}{2}{0mm}{-1\baselineskip}{1\baselineskip}{\normalfont\large\bfseries}}
\renewcommand{\subsection}{\@startsection{subsection}{2}{0mm}{-0.8\baselineskip}{0.6\baselineskip}{\normalfont\normalsize\bfseries}}
\renewcommand{\subsubsection}{\@startsection{subsubsection}{3}{0mm}{-0.6\baselineskip}{0.4\baselineskip}{\normalfont\normalsize\itshape}}
\begin{document}

\begin{bibunit}

\title{Difference-in-Differences Estimators When No Unit Remains Untreated\thanks{This paper was previously circulated as ``An Optimal Bandwidth For Difference-in-Difference Estimation with a Continuous Treatment and an Heterogeneous Adoption Design'' and ``Two-way Fixed Effects and Difference-in-Difference Estimators in Heterogeneous Adoption Designs without Stayers''. We are very grateful to Marc Gurgand for a discussion that initiated this paper, and  to Marc Kaufmann, Tomasz Olma, Roland Rathelot and Adam Szeidl for their helpful comments. Cl\'{e}ment de Chaisemartin was funded by the European Union (ERC, REALLYCREDIBLE,GA N°101043899). Views and opinions expressed are those of the authors and do not reflect those of the European Union or the European Research Council Executive Agency. Neither the European Union nor the granting authority can be held responsible for them. \textbf{AI disclosure statement}: ChatGPT helped us to rewrite the paper, it found and corrected an error in a result discussed in Section \ref{ssub:noQUG}, and it found some typos in the proof of Theorem \ref{thm:test_inf_support}.}}

\author{Cl\'{e}ment de Chaisemartin\thanks{Sciences Po Paris, clement.dechaisemartin@sciencespo.fr} \and Diego Ciccia\thanks{Northwestern University, Kellogg School of Management, diego.ciccia@kellogg.northwestern.edu} \and Xavier D'Haultf\oe{}uille\thanks{CREST-ENSAE, xavier.dhaultfoeuille@ensae.fr} \and Felix Knau\thanks{Ludwig-Maximilians-University Munich, felix.knau@econ.lmu.de}}

\maketitle

\begin{abstract}
We study difference-in-differences designs in which all units are initially untreated and later receive heterogeneous, strictly positive treatment doses. When quasi-untreated units receive doses close to zero, a difference-in-difference estimand using those units as the control group identifies a weighted average treatment effect per unit of treatment, under a parallel-trends assumption. We develop a nonparametric estimator and bias-corrected confidence intervals for this parameter. We also show that the assumptions justifying the usual two-way fixed-effects estimator imply testable linearity restrictions, and propose specification tests that can guide its use.
\end{abstract}
\newpage

\section{Introduction}

We consider treatment-effect estimation in designs in which no unit is treated initially, and then units simultaneously receive heterogeneous and strictly positive treatment doses. We refer to such designs as heterogeneous adoption designs (HADs). To simplify the exposition, in the introduction we focus on the case with two time periods: units are untreated at period one and treated at period two. In HADs, a common estimation strategy is to run a two-way fixed effects (TWFE) regression of $Y_{g,t}$, the outcome of unit $g$ at period $t$, on unit fixed effects, an indicator for period two, and $D_{g,t}$, the treatment dose received by $g$ at $t$. Proposition S1 of \cite{dcDH2020} considers TWFE regressions in HADs, and shows that
under a parallel-trends assumption, they fail to identify a convex combination of unit-specific effects \citep[the same result also appears in][]{callaway2021difference}. Then, they could be misleading if effects vary across units. Another common estimation strategy is to regress $Y_{g,t}$, on unit fixed effects, an indicator for period two, and an indicator for receiving a treatment dose above, say, the median. This strategy can also be misleading if effects vary across units \citep{fricke2017identification}.

\medskip
This lack of robustness matters because HADs are common. They arise when a policy is implemented universally, but exposure varies. For instance, nationwide increases in the minimum wage affect industries or local areas differentially, depending on their prior fraction of workers below the new minimum wage \citep[see, e.g.,][]{dustmann_reallocation_2022}. Similarly, China's accession to the WTO eliminated uncertainty about US--China trade tariffs, but affected US industries differentially depending on their prior levels of tariff uncertainty \citep{pierce2016surprisingly}. Likewise, the creation of Medicare Part D affected drugs differentially according to their Medicare market share \citep{duggan2010}. Such designs also arise when a new medical treatment is introduced but its prescription rate varies across hospitals \citep{Chaisemartin2011fuzzy,melamed2021association}. In all these examples, no unit remains untreated: all industries or local areas have some workers earning less than the new nationwide minimum wage, every drug had a nonzero Medicare market share prior to the creation of Medicare Part D, etc. Even when untreated units exist, researchers may prefer to discard them, for fear they differ too much from treated units.

\medskip
To address the lack of robustness of the TWFE estimator, this paper proposes two approaches: a heterogeneity-robust DID approach using quasi-untreated units as controls, and a specification test for the assumptions justifying the TWFE estimator. Both approaches are complementary: in view of its slow rate of convergence the nonparametric estimator will often be less precise than the TWFE estimator, so the TWFE estimator may still be appealing when its assumptions are not rejected.

\subsection*{A nonparametric estimator in designs with quasi-untreated units}

In the first part of the paper, we assume that there is a quasi-untreated group (QUG), that has a period-two treatment local to zero. Then, we show that under a parallel trends assumption, the average effect of actual treatment exposure normalized by the average treatment dose, a parameter we call the weighted average slope (WAS), is identified by an estimand comparing the outcome evolutions of the whole population and of the QUG. In the same spirit as familiar regression discontinuity estimators \citep[see][]{Imbens2012,calonico2014}, we leverage results from the nonparametric estimation literature \citep[see][]{calonico2018effect} to propose: (i) an optimal bandwidth defining the QUG in a data-driven way, (ii) an estimator relying on a local-linear regression, and (iii) a robust confidence interval. Our estimator converges at the standard univariate nonparametric rate. This estimator and its confidence interval are computed by the \st{did\_had} Stata, R, and Python packages. We also propose a nonparametric and tuning-parameter-free test of the null hypothesis that there is a QUG.

\subsection*{A test of the assumptions underlying the TWFE estimator}

As mentioned above, the TWFE estimator is not robust to heterogeneous treatment effects in HADs. But it does identify a causal effect if, on top of parallel trends, one also imposes an homogeneous-and-linear effect condition.
We show that those two assumptions imply
\begin{equation}\label{eq:lin_intro}
E(\Delta Y_g|D_{g,2})=\beta_0+\beta_{fe} D_{g,2}:
\end{equation}
i.e. the conditional mean of the outcome's evolution should be linear in the period-two dose. Therefore, curvature in the conditional mean directly falsifies the joint assumptions justifying TWFE. Under parallel-trends, we show that in designs with a QUG, there is actually an ``if and only if'' between \eqref{eq:lin_intro} and the homogeneous and linear effect assumption. This motivates the following recommendations: if a linearity test of $E(\Delta Y_g|D_{g,2})$ rejects, do not use the TWFE estimator; if a linearity test of \(E(\Delta Y_g\mid D_{g,2})\), a pre-trend test of parallel trends, and our test for the presence of a QUG do not reject and have decent power, one may use the TWFE estimator; if the first two tests do not reject but the third does (there is no QUG), one may use the TWFE estimator, acknowledging one could only test an implication of its assumptions. Testing \eqref{eq:lin_intro} is straightforward when $D_{g,2}$ takes a finite number of values. When $D_{g,2}$ is continuous, we use the nonparametric and tuning-parameter free test proposed by \cite{stute1997nonparametric} and \cite{stute1998bootstrap}, which we implemented in the \st{stute\_test} Stata and R packages.

\subsection*{Application}

We use our results to revisit \cite{garrett_tax_2020}. In the 2002 Job Creation and Worker Assistance Act, the US government introduced a bonus depreciation regulation, whereby firms can deduct 30 percent of the purchase price of a new investment from their taxable income, yielding a decrease in the present value of investment costs. The authors define a county-level treatment capturing exposure to that reform. Our nonparametric WAS estimate indicates a positive and significant treatment effect on employment, and we find little evidence of differential pre-trends. The TWFE estimate is twice smaller, and the Stute linearity test is very strongly rejected. Thus, the WAS estimate seems more reliable in this application.

\subsection*{Related literature and organization of the paper}

Few papers consider continuous-treatment designs in which comparisons must be made across treatment doses. \cite{fricke2017identification} studies DIDs comparing more- and less-treated groups and shows that interpreting such comparisons as causal effects of moving between treatment levels generally requires restrictions on treatment-effect heterogeneity. \cite{callaway2021difference} develop a broader framework for DIDs with continuous treatments. They show that with untreated units, under a standard parallel-trends assumption, treatment-on-the-treated effects at each dose can be identified, but comparisons of these effects across doses generally combine causal responses with selection arising from treatment-effect heterogeneity. They also discuss designs without untreated units, where differences in ATT-type effects across doses remain identified under standard parallel trends, but interpreting those differences as causal responses still requires stronger restrictions that rule out selection arising from treatment-effect heterogeneity. By contrast, our nonparametric estimator identifies an average treatment effect per unit of treatment using quasi-untreated units, without imposing restrictions on treatment-effect heterogeneity across dose groups.

\medskip
Many HADs, including the minimum-wage and medical-treatment examples above, are so-called \textit{fuzzy designs} \citep{deChaisemartin15b}, where a unit's treatment dose equals the fraction of individuals within the unit who are treated. When $D_{g,1}=0$, as we assume here, Point 2 of Theorem 1 of \cite{deChaisemartin15b} implies that a DID estimand using units untreated at period two as the control group only relies on a parallel-trends assumption, an identification result similar to our Theorem 1 below. However, they do not propose an estimator with proven guarantees when the proportion of units untreated at period two is equal to zero, the case we consider.

\medskip
Our paper is also related to the literature that considers correlated random coefficient (CRC) estimators. Those estimators rely on a strong exogeneity assumption similar to parallel trends, and they allow for heterogeneous treatment effects. However, we are not aware of a CRC estimator applicable to HADs.\footnote{The most closely related paper is \cite{graham2012identification}, who, unlike \cite{chamberlain1992efficiency}, allow for designs with two time periods. Similar to our paper, they propose to use quasi stayers as the control group. However their estimator relies on the assumption that $D_{g,2}-D_{g,1}$ can take both positive and negative values, while $D_{g,2}-D_{g,1}>0$ in an HAD. Specifically, if $T=p=2$, $D_{g,1}=0$, and $D_{g,2}>0$, then in their Theorem 2.1, the asymptotic variance of the second coordinate of $\widehat{\beta}$, the average of the random slopes, is equal to  $+\infty$. Instead, our estimator, which leverages modern local-linear regression estimators, can be used even if $D_{g,2}-D_{g,1}>0$.}

\medskip
The paper is organized as follows. Section \ref{sec:setup} introduces the set-up. Section \ref{sec:HR} introduces our nonparametric estimator. Section \ref{sec:TWFE} considers the TWFE estimator. Section \ref{sec:appli} revisits an empirical application. Proofs are collected in the appendix.

\section{Setup}\label{sec:setup}

\paragraph{Panel data.} We are interested in estimating the effect of a treatment on an outcome, using a panel data set, with $G$ units and two time periods. In Section \ref{sub:extensions}, we discuss how results generalize to designs with more time periods. Units could be aggregate entities, like regions or industries.  Henceforth, units are indexed by $g$ and time periods by $t$.

\paragraph{Treatment and potential outcomes.}
Let $D_{g,t}$ denote the treatment of unit $g$ at $t$, with $D_{g,t}\ge 0$. The potential outcome of unit $g$ at $t$ under treatment $d$ is $Y_{g,t}(d)$, and the observed outcome is $Y_{g,t}:=Y_{g,t}(D_{g,t})$. Our potential outcome notation rules out anticipatory effects: units' outcome at period one does not depend on their period-two treatment. Our notation also rules out dynamic or carry-over effects:  units' outcome at period two does not depend on their past treatments. In the designs we consider, units are all untreated at period one and at prior periods, because the treatment does not exist before $t=2$. Then, the no carry-over effects assumption is not of essence: $Y_{g,2}(0)$ simply stands for the potential outcome of $g$ at period $2$ if it has never been treated, while $Y_{g,2}(d)$ stands for its potential outcome if it has never been treated till period $1$ and then receives dose $d$ at period $2$. On the other hand, the no-anticipation assumption is of essence.

\paragraph{Notation and convention.}
Hereafter, we let $\Supp(A)$ denote the support of the random variable $A$, namely the smallest closed set $C$ such that $P(A\in C )=1$. Throughout the paper, whenever a limit is introduced in an assumption, it is implicitly assumed that this limit exists. $\Delta$ denotes the first-difference operator.

\subsection{Assumptions}

\begin{hyp} (i.i.d. sample)
    $(Y_{g,1},Y_{g,2},D_{g,1},D_{g,2})_{g=1,...,G}$ are i.i.d.
    \label{hyp:iid}
\end{hyp}
As units are i.i.d., we drop the $g$ subscript below, except when we introduce estimators. Let $\underline{d}:=\inf \Supp(D_2)$ be the dose received by the least-treated units.
\begin{hyp}
	(The least treated and the full population are on parallel trends)\\
    $\lim_{d\downarrow \underline{d}}E\left[\Delta Y(0)|D_2\le d\right]=E\left[\Delta Y(0)\right]$.
	\label{hyp:PT_least}
\end{hyp}
Assumption \ref{hyp:PT_least} requires that the least-treated experience the same average outcome evolution without treatment as the overall population. If the data contains another pre-treatment period $t=0$, we can assess the plausibility of Assumption \ref{hyp:PT_least} by computing a pre-trend estimator, see Section \ref{sub:extensions}. With a binary treatment, Assumption \ref{hyp:PT_least} reduces to the standard parallel-trends assumption: it requires that untreated and treated units have the same average outcome evolutions without treatment.
\begin{hyp}
	(Bounded treatment effects) There exists $M>0$ such that for all $(d,d')\in \Supp(D_2)^2$, $|(Y_2(d)-Y_2(d'))/(d-d')|\le M $.
\label{hyp:bounded_slopes}
\end{hyp}
%{\bf XDH I would rather use: \begin{hyp}
%	(Uniform continuity of potential outcomes) There exists $d_0>\underline{d}$ and a continuous function $M$ such that $M(0)=0$ and for all $(d,d')\in [\underline{d},d_0]^2$, $|Y_2(d)-Y_2(d')|\le M(d-d')$.
%	\label{hyp:bounded_slopes}
%\end{hyp}}
Assumption \ref{hyp:bounded_slopes} requires that the treatment effect is bounded by a (potentially very large) constant $M$.\footnote{Assumption \ref{hyp:bounded_slopes} is stronger than needed for our results to hold, it could be replaced by a uniform continuity condition.}
When $\underline{d}=0$, Assumption \ref{hyp:bounded_slopes} implies that $d\mapsto Y_2(d)$ is continuous at $0$. This rules out extensive margin effects, where infinitesimally small treatment doses already have a non-zero effect \citep[see, e.g.,][who also rules out extensive-margin effects in a different context]{caetano2015test}.

%Assumption \ref{hyp:bounded_slopes} is a regularity condition,  which requires that the slopes of the period-two potential outcome be bounded. It automatically holds if $d\mapsto Y_2(d)$ is almost surely differentiable for all $d$ with a bounded derivative. Importantly, we do not assume that the researcher knows the value of $M$.

%Actually, we could consider an even weaker condition, by assuming that the limit of $d\mapsto (d-\underline{d}) |E\left[\text{TE}_2|D_2\le d\right]|$ at $\underline{d}$ is zero. Nonetheless, when $\underline{d}=0$, both Assumption \ref{hyp:bounded_slopes} and this weaker version  rule out extensive effects, in which $d\mapsto Y_2(d)$ would be discontinuous at $0$.
%Assumption \ref{hyp:bounded_slopes} requires that the slopes conditional on the period-two treatment, $E\left[\text{TE}_2|D_2\right]$, be bounded. This is a weak condition: for instance, it automatically holds if $d\mapsto Y_2(d)$ is almost surely differentiable with a bounded derivative. Actually, we could consider an even weaker condition, by assuming that the limit of $d\mapsto (d-\underline{d}) |E\left[\text{TE}_2|D_2\le d\right]|$ at $\underline{d}$ is zero. Nonetheless, when $\underline{d}=0$, both Assumption \ref{hyp:bounded_slopes} and this weaker version  rule out extensive effects, in which $d\mapsto Y_2(d)$ would be discontinuous at $0$.

\subsection{Heterogeneous adoption designs}\label{subsec:HAD}

We consider heterogeneous adoption designs (HADs), where all units are untreated at period one and receive heterogeneous, strictly positive treatment doses at period two. %Moved below Our results also apply to designs where units all receive the same non-zero treatment dose $d_1\ne 0$ at period one. On the other hand, they do not easily generalize to designs where units receive heterogeneous doses at periods one \citep{de2024difference}. Then, at period two units receive heterogeneous, strictly positive treatment doses.
\begin{des}\label{des:HAD} (HAD)
$D_{1}=0$, $D_{2}>0$, and $V(D_2)>0$.
\end{des}
Our results also apply to designs where units all receive the same non-zero treatment dose $d_1\ne 0$ at period one. On the other hand, they do not easily generalize to designs where units receive heterogeneous doses at periods one \citep{de2024difference}.

\medskip
Some of our results apply to a subset of Design \ref{des:HAD}, namely designs with a QUG.
\setcounter{des}{0}
\begin{des}\hspace{-0.22cm}{\bf '} \label{des:HAD_stayersorquasistayers}
(HAD with a quasi-untreated group)
The conditions in Design \ref{des:HAD} hold and $\underline{d}=0$.
\end{des}
The condition in Design \ref{des:HAD_stayersorquasistayers}' holds when there are units whose period-two treatment is ``very close'' to zero: for any $\delta>0$, $P(0<D_2<\delta)>0$. For instance, it holds if $D_2$ is continuously distributed on $\mathbb{R}_+$ with a continuous density that is strictly positive at 0.

\section{Nonparametric estimator with a quasi-untreated group}
\label{sec:HR}

Throughout this section, we assume that $\underline{d}=0$: we are in a design with a QUG.

\subsection{Target parameter}\label{subsec:target}

In the non-binary treatment setting we consider, a natural target is $$d\mapsto E\left(\frac{Y_2(d)-Y_2(0)}{d}\right),$$ which may for instance be used to assess if the treatment has increasing, decreasing, or constant returns. However, for $0<d\ne D_2$, estimating $(Y_2(d)-Y_2(0))/d$ requires estimating both $Y_2(0)$ and $Y_2(d)$. As $Y_2(d)$ is not observed at $t=2$ and at prior periods, estimating it requires making assumptions whose plausibility cannot be assessed via pre-trend tests, while tests or placebo tests of identifying assumptions have become a hard-to-bypass standard in observational research \citep{imbens2024lalonde}. Instead, we focus on
\begin{align*}
\text{WAS}:=& E\left[\frac{D_2}{E[D_2]}\text{TE}_2\right],
\end{align*}
a weighted average of
$$\text{TE}_2:=\frac{Y_2(D_2)-Y_2(0)}{D_2},$$
the slopes of units' potential outcome functions between $0$ and their actual treatments, where units with a larger period-two treatment receive more weight.
Estimating $\text{TE}_2$ only requires estimating  $Y_2(0)$, which can sometimes be achieved under Assumption \ref{hyp:PT_least}, a parallel-trends assumption whose plausibility can be assessed via pre-trend tests. As
\begin{align}\label{eq:HAD_WaldDID}
\text{WAS}=&E\left[\frac{D_2}{E[D_2]}\frac{Y_2(D_2)-Y_2(0)}{D_2}\right]\nonumber\\
=&\frac{E[Y_2(D_2)-Y_2(0)]}{E[D_2]}\nonumber\\
=&\frac{E[Y_2(D_2)-Y_2(0)|D_2>0]}{E[D_2|D_2>0]},
\end{align}
WAS is just equal to $E[Y_2(D_2)-Y_2(0)|D_2>0]$, a generalization of the standard average treatment effect on the treated to designs with a non-binary treatment, normalized by $E[D_2|D_2>0]$, the average treatment dose received by treated units.

\subsection{Identification}
\label{sub:identification_WAS}

\begin{thm}\label{thm:HAD_WaldDID}
Suppose that we are in Design \ref{des:HAD_stayersorquasistayers}' and Assumptions \ref{hyp:PT_least} and \ref{hyp:bounded_slopes} hold. Then,
\begin{equation}\label{eq:HAD_thm_WaldDID}
\text{WAS}=\frac{E[\Delta Y]-\lim_{d\downarrow 0} E\left[\Delta Y|D_2\le d\right]}{E[D_2]}.
\end{equation}
\end{thm}
Theorem \ref{thm:HAD_WaldDID} shows that with a QUG, $\text{WAS}$ is identified by an estimand comparing the average outcome evolution to that of the QUG, scaled by the average treatment.

\subsection{Estimation and inference}
\label{sub:estimation_WAS}

To estimate $\lim_{d\downarrow 0} E\left[\Delta Y|D_2\le d\right]$, we rely on a local linear regression. Let
$$\widehat{\beta}^{\text{np}}_h:=\frac{\frac{1}{G}\sum_{g=1}^G \Delta Y_g - \widehat{\mu}_h}{\frac{1}{G}\sum_{g=1}^G D_{g,2}},$$
    with $\widehat{\mu}_h$ the intercept in the local linear regression of $\Delta Y_g$ on $D_{g,2}$, weighting observations by $k(D_{g,2}/h)/h$, for a kernel function $k$ and a bandwidth $h>0$. One can derive a so-called optimal bandwidth $h_G^*$, which, as in RDDs \citep[see][]{Imbens2012}, minimizes the asymptotic mean squared error of $\widehat{\beta}^{\text{np}}_h$. Then, inference on $\text{WAS}$ is not straightforward,
because the asymptotic distribution of
$$\sqrt{Gh_G^*}(\widehat{\beta}^{\text{np}}_{h_G^*}-\text{WAS})$$
has a first-order bias. However, the approach for local-polynomial regressions in \cite{calonico2018effect} readily applies to our set up. We rely on their results and software  implementation \citep[see][]{calonico2019nprobust} to:
\begin{enumerate}
\item estimate a mean-squared error (MSE) optimal bandwidth $\widehat{h}^*_G$;
\item compute $\widehat{\mu}_{\widehat{h}^*_G}$;
\item compute $\widehat{\mathcal{B}}_{\widehat{h}_G^*}$, an estimator of $\widehat{\mu}_{\widehat{h}^*_G}$'s first-order bias;
\item compute $\widehat{V}_{\widehat{h}_G^*}$, an estimator of the variance of $\widehat{\mu}_{\widehat{h}^*_G}-\widehat{\mathcal{B}}_{\widehat{h}_G^*}$.
\end{enumerate}
Then, we define our estimator as
\begin{equation}\label{eq:beta_QS}
\widehat{\beta}^{\text{np}}_{\widehat{h}_G^*}:=\frac{\frac{1}{G}\sum_{g=1}^G \Delta Y_g - \widehat{\mu}_{\widehat{h}^*_G}}{\frac{1}{G}\sum_{g=1}^G D_{g,2}},
\end{equation}
and its bias-corrected confidence interval as
\begin{equation}\label{eq:CI_QS}
\left[\widehat{\beta}^{\text{np}}_{\widehat{h}_G^*}+\frac{\widehat{\mathcal{B}}_{\widehat{h}_G^*}}{\frac{1}{G}\sum_{g=1}^G D_{g,2}} \pm\frac{q_{1-\alpha/2}\sqrt{\widehat{V}_{\widehat{h}_G^*}/(G\widehat{h}_G^*)}}{\frac{1}{G}\sum_{g=1}^G D_{g,2}}\right],
\end{equation}
where $q_x$ denotes the quantile of order $x$ of a standard normal. This confidence interval ignores the variation comming from $G^{-1}\sum_{g = 1}^G \Delta Y_g$ and $G^{-1}\sum_{g = 1}^G \Delta D_g$, as they converge at a faster rate than $\widehat{\mu}_{\widehat{h}^*_G}$. If Assumptions \ref{hyp:iid}-\ref{hyp:bounded_slopes} and \ref{hyp:regul} in the appendix hold, Theorem 2(c) of \cite{calonico2018effect}, %Suppose that Assumptions \ref{hyp:iid}-\ref{hyp:bounded_slopes} and \ref{hyp:regul} in the appendix hold. Then Theorem 2(c) of \cite{calonico2018effect},
applied with $p=1$ and $q=2$ at the boundary point zero, implies that the confidence interval in \eqref{eq:CI_QS} has asymptotic coverage $1-\alpha$.
Moreover, since the MSE-optimal bandwidth is of order \(G^{-1/5}\), \(\widehat\beta^{np}_{\widehat h_G^*}\) converges to WAS at the rate $G^{-2/5}$. %Lastly, note that $G^{-1}\sum_{g = 1}^G \Delta Y_g$ is $G^{1/2}$-consistent for $E[\Delta Y]$. As a result, estimation error for the first term of the numerator of $\widehat\beta^{np}_{\widehat h_G^*}$ is asymptotically negligible.

\medskip
Assumption \ref{hyp:regul} requires that the density of $D_2$ is strictly positive at zero. As $0$ lies on the boundary of the support of $D_2$, this assumption is not innocuous. Proposing an inference method for the case where that density is zero is an interesting avenue for future research, which goes beyond the scope of this paper.\footnote{\cite{hengartner1996nonparametric} study the behavior of local linear estimators at points with zero density and show that they achieve slower rates of convergence than when the density is strictly positive. However, to our knowledge no inference method has been developed yet in such cases. Ideally, one would even want an approach that adapts to whether the density is zero or not, a challenging problem.}
In Appendix \ref{sub:sims}, we conduct simulations to assess the coverage of the CI in \eqref{eq:CI_QS}. Its coverage is close to nominal for moderately large sample sizes ($G\geq 500$) even in DGPs where Assumption \ref{hyp:regul} fails because the density of $D_2$ is equal to zero at zero.

\subsection{Pre-trend and event-study estimators}\label{sub:extensions}

In this section, we no longer assume that we have only two time periods. Instead, we have $T>2$ periods, and we make the following assumption on the design:
\begin{des}\label{des:HAD_multipleT} (HAD with multiple periods)
There exists $F\geq 2$ such that $D_{t}=0$ for all $t<F$, $D_{F}>0$, $V(D_F)>0$, and $D_t=D_F$ for all $t>F$.
\end{des}
$F$ represents the period when treatment starts. Units are all untreated before $F$, all start being treated at $F$, and all keep the same treatment after $F$. These conditions for instance hold in \cite{dustmann_reallocation_2022}, \cite{pierce2016surprisingly}, and \cite{duggan2010}, some of the examples we mentioned in the introduction. They also hold in \cite{garrett_tax_2020}, the empirical example we revisit below.

\paragraph{Pre-trend estimators.}  If $F>2$, for all $t<F-1$ we can define a pre-trend estimator mimicking $\widehat{\beta}^{\text{np}}_{\widehat{h}_G^*}$, where we just replace $\Delta Y_g$ by $ Y_{g,t}-Y_{g,F-1}$ and $D_{g,2}$ by $D_{g,F}$ in the estimator's definition. One can show that if the analogue of Assumption \ref{hyp:PT_least} holds from period 1 to $F-1$, these pre-trend estimators converge to zero, so this assumption is testable.

\paragraph{Event-study estimators.}
Similarly, for all $t\geq F$ we can define an event-study estimator mimicking $\widehat{\beta}^{\text{np}}_{\widehat{h}_G^*}$, where one replaces $\Delta Y_g$ by $ Y_{g,t}-Y_{g,F-1}$ and $D_{g,2}$ by $D_{g,F}$ in the estimator's definition. Letting $\bm{0}_{k}$ and $\bm{D}_{F,k}$ denote $1\times k$ vectors with all coordinates equal to $0$ and $D_F$, respectively, one can show that if the analogue of Assumption \ref{hyp:PT_least} holds from period $F-1$ to $T$, this estimator estimates a weighted average of $$(Y_{t}(\bm{0}_{F-1},\bm{D}_{F,t-(F-1)})-Y_{t}(\bm{0}_{t}))/D_{F},$$ the effect of having received a dose of $D_{F}$ rather than $0$ for $t-(F-1)$ periods, normalized by $D_{F}$.\footnote{In the previous display, the potential outcome notation explicitly allows for carry-over effects.} Finally, if units' treatment can keep changing after period $F$ ($D_t\ne D_F$ for $t>F$), we recommend computing  an event-study estimator mimicking $\widehat{\beta}^{\text{np}}_{\widehat{h}_G^*}$, where one replaces $\Delta Y_g$ by $ Y_{g,t}-Y_{g,F-1}$ and $D_{g,2}$ by $D_{g,t}$. Under the assumption that lagged treatments do not affect the outcome (no carry-over effect), this estimator estimates a weighted average of $(Y_{t}(D_t)-Y_{t}(0))/D_t.$
Proposing estimators that allow for carry-over effects of lagged treatments in HADs where groups' treatment can change more than once is an interesting avenue for future research.
%If $D_{g,3}\ne D_{g,2}$, meaning that units' treament can change from period two to three, on top of replacing $\Delta Y_g$ by $ Y_{g,3}-Y_{g,1}$, one should also replace $D_{g,2}$ either by $D_{g,3}$ if one is ready to rule out carry-over effects, or by $D_{g,2}+D_{g,3}$ See Appendix \ref{sec:extension_multipleT} for further details on the extension to multiple periods.

\subsection{Computation}\label{sub:computation}

The \st{did\_had} Stata, R, and Python \citep[see][]{did_had} commands compute $\widehat{\beta}^{\text{np}}_{\widehat{h}_G^*}$ and its bias-corrected confidence interval. When more than one pre-treatment period is available, the commands can compute pre-trend estimators. With more than one post-treatment period, the commands can compute event-study estimators.
%The basic syntax of the Stata command is: \medskip
%\st{did\_had Y G T D }\\
\st{did\_had} relies on the \st{nprobust} package of \cite{calonico2019nprobust}.

\subsection{Testing the null that there is a QUG}\label{sub:testQS}

This section assumes there is a QUG: $\underline{d}=0$. We now propose a simple and tuning-parameter-free test of the null hypothesis that $\underline{d}=0$. The test statistic is $T = D_{2,(1)}/(D_{2,(2)}-D_{2,(1)})$, where $D_{2,(1)}\le ... \le D_{2,(G)}$ denotes the order statistic of $(D_{2,g})_{g=1,...,G}$ ($T=\infty$ if $D_{2,(2)}=D_{2,(1)}$). For an $\alpha$-level test, the critical region is $W_\alpha:=\{T>1/\alpha - 1\}$. Intuitively, we reject the null if the distance between $D_{2,(1)}$ and 0 is more than $1/\alpha - 1$ times larger than the distance between $D_{2,(2)}$ and $D_{2,(1)}$. We show that this test has asymptotic size equal to $\alpha$ and nontrivial local power on a broad class of cdfs for $D_2$. Specifically, let $\mathcal{D}$ denote the set of cdfs on the real line, let $\overline{d}>\underline{d}\ge 0$, $m>0$ and $K>0$, and let
\begin{align*}
\mathcal{F}^{\underline{d}, \overline{d}}_{m,K} := & \; \bigg\{F\in \mathcal{D}: F \text{ is differentiable on } [\underline{d}, \overline{d}], \; F(\underline{d}) =0, \, F'(d)\ge m\; \forall d\in [\underline{d}, \overline{d}] \\ & \; \text{ and } |F'(d) -F'(d_1)|\le K |d-d_1| \; \forall (d_1,d)\in [\underline{d}, \overline{d}]^2\bigg\}.	
\end{align*}
This set includes differentiable cdfs whose support has infimum equal to $\underline{d}$ and with density bounded from below in a neighborhood of $\underline{d}$. Hereafter, we denote probabilities by $P_F$ instead of $P$, to emphasize their dependence on the cdf of $D_2$.

\begin{thm}\label{thm:test_inf_support} Fix $\alpha\in (0,1)$, $\overline{d}>\underline{d}> 0$, $m\in (0,1/\overline{d}]$, and $K>0$. We have:
\begin{enumerate}
	\item (Asymptotic size control) $\lim\sup_{G\to\infty} \sup_{F\in \mathcal{F}^{0,\overline{d}}_{m,K}} P_F(W_\alpha) = \alpha$.
	\item (Uniform consistency) $\lim\inf_{G\to\infty} \inf_{F\in \mathcal{F}^{\underline{d},\overline{d}}_{m,K}} P_F(W_\alpha) = 1$.
	\item (Local power) $\forall (\underline{d}_G)_{G\ge 1}: d_G \to 0$ and $\lim\inf G \underline{d}_G =c >0$, $$\lim\inf_{G\to\infty} \inf_{F\in \mathcal{F}^{\underline{d}_G,\overline{d}}_{m,K}} P_F(W_\alpha) \ge 1-(1-\alpha)\exp\left(-\frac{\alpha cm}{1-\alpha}\right)  >\alpha.$$
\end{enumerate}
\end{thm}
Point 1 of Theorem \ref{thm:test_inf_support} establishes the asymptotic validity of the test. Point 2 shows that the test is consistent against fixed alternatives. Finally, Point 3 shows that the test has power against local alternatives converging to the null at the rate $1/G$.

\medskip
We build on the classical theory of extreme order-statistic spacings \citep{pyke1965spacings}, which provides the limiting distribution underlying our test statistic. Our test is also related to the test of limited overlap in \cite{ma2025testing}. The two tests assess different null hypotheses: their test assesses if the expectation of the inverse propensity score is bounded, whereas ours assesses whether zero is the lower endpoint of the treatment-dose support. While the two test statistics are different, both procedures use a fixed number of extreme order statistics and normalize by extreme spacings, avoiding consistent estimation of a tail index.

\medskip
We conjecture that under the null that $\underline{d}=0$, drawing inference on WAS only conditional on not rejecting the pre-test above does not distort inference. The reason behind this conjecture is asymptotic independence of extreme-order statistics and sample averages, see e.g. Theorem 2.4 in \cite{li2024limit}. Also, $T$ above only depends on the extremes $(D_{2,(1)}, D_{2,(2)})$ and  \cite{calonico2014} show that up to a remainder term, $\widehat{\mu}_{\widehat{h}^*_G}-\widehat{\mathcal{B}}_{\widehat{h}_G^*}$ is equal to the sample mean of some appropriate i.i.d. variables. However, the distribution of these variables depends on $G$, and we are not aware of a generalization of  Theorem 2.4 in \cite{li2024limit} to triangular arrays.

\subsection{What can be learned without quasi-stayers?}\label{ssub:noQUG}

In previous versions of this paper \citep[see][]{dechaisemartin2024twowayfixedeffectsdifferencesindifferencesv4,dechaisemartin2026differenceindifferencesestimatorsunitremains}, we proposed nonparametric identification strategies in designs without quasi-stayers.

\medskip
A first possibility is to restrict treatment effect heterogeneity, by assuming that
\begin{equation}\label{eq:hyp_sign_identification}
|\lim_{d\downarrow \underline{d}} E(\text{TE}_2|D_2\le d)|< \frac{E(D_2)}{\underline{d}}|\text{WAS}|:
\end{equation}
the effect among the least treated is bounded, in absolute value, by the absolute value of the WAS multiplied by the ratio of the average and lowest treatment dose. This ratio can be estimated, it is necessarily greater than one, and it may often be much greater than one.\footnote{Assumption 5 in \cite{dechaisemartin2026differenceindifferencesestimatorsunitremains} has an error, its left hand side should have an absolute value; this error is corrected in \eqref{eq:hyp_sign_identification}.} Under Assumption \ref{hyp:PT_least},
\[
E[\Delta Y]-\lim_{d\downarrow\underline d}E[\Delta Y\mid D_2\le d]
=
E[D_2]\mathrm{WAS}
-\underline d\lim_{d\downarrow\underline d}E[\mathrm{TE}_2\mid D_2\le d].
\]
If one also imposes \eqref{eq:hyp_sign_identification},
the previous display implies that $E[\Delta Y]-\lim_{d\downarrow\underline d}E[\Delta Y\mid D_2\le d]$ and WAS are of the same sign, so the sign of WAS is identified.

\medskip
A second possibility, in the spirit of \cite{fricke2017identification} and \cite{callaway2021difference}, is to assume that the average effect of receiving the lowest treatment dose is the same among the least treated and in the full population:
\begin{equation}\label{eq:hyp_id_withoutQS}
\lim_{d\downarrow \underline{d}} E\left[Y_2(\underline{d})- Y_2(0)|D_2\le d\right]=E\left[Y_2(\underline{d})- Y_2(0)\right].
\end{equation}
Under Assumption \ref{hyp:PT_least} and \eqref{eq:hyp_id_withoutQS}, one can show that
$$\frac{E\left[Y_2(D_2)-Y_2(\underline{d}) \right]}{E[D_2-\underline{d}]}=\frac{E[\Delta Y]-\lim_{d\downarrow \underline{d}} E\left[\Delta Y |D_2\le d\right]}{E[D_2-\underline{d}]}:$$
the effect of receiving a dose of $D_2$ rather than $\underline{d}$, scaled by the average of $D_2-\underline{d}$, is identified by an estimand that generalizes that in Theorem \ref{thm:HAD_WaldDID}, where $0$ is replaced by the lowest treatment dose $\underline{d}$. Thus, $E\left[Y_2(D_2)-Y_2(\underline{d}) \right]/E[D_2-\underline{d}]$ can be estimated nonparametrically.

\medskip
Note that both strategies replace the QUG with restrictions on treatment-effect heterogeneity whose plausibility cannot be assessed using pre-trend tests and which, under the maintained assumptions alone, are not testable.

\section{TWFE estimator}\label{sec:TWFE}

\paragraph{Motivation.}
When there is a QUG, we have proposed a $G^{2/5}$-consistent treatment-effect estimator, and $G^{2/5}$-consistent pre-trend estimators that one can use to assess the plausibility of the parallel-trends assumption underlying the estimator. Yet, in view of its slow rate of convergence, when the sample size is not very large that estimator may have limited power to detect the treatment effect and differential trends. Moreover, these estimators can only be used with a QUG. This leads us to reconsider the usual TWFE estimator in this section.

\paragraph{Definition of the TWFE estimator.}
Let $\widehat{\beta}_{fe}$ denote the coefficient on $D_{g,2}$ in a regression of $\Delta Y_g$ on a constant and $D_{g,2}$. $\widehat{\beta}_{fe}$ is equal to the coefficient on  $D_{g,t}$ in a regression of $Y_{g,t}$ on unit fixed effects, an indicator for period $2$, and $D_{g,t}$. Let
$$\beta_{fe}:=\frac{E[(D_{2}-E(D_{2}))\Delta Y]}{E[(D_{2}-E(D_{2}))D_{2}]}$$
denote the probability limit of $\widehat{\beta}_{fe}$ when $G\rightarrow +\infty$ under Assumption \ref{hyp:iid} and if $E[D_2^2]<\infty$ and $E[\Delta Y^2]<\infty$.

\subsection{Conditions under which $\widehat{\beta}_{fe}$ is consistent for a causal effect.}

\begin{hyp}
	(Parallel trends across all treatment doses) There is a real number $\mu_0$ such that $E\left[\Delta Y(0)|D_2\right]=\mu_0$.
	\label{hyp:strong_exogeneity}
\end{hyp}
Assumption \ref{hyp:strong_exogeneity} requires that units receiving different treatment doses at period two would all have experienced parallel trends without treatment. This assumption, which is similar to a strong exogeneity assumption in panel data models, is stronger than Assumption \ref{hyp:PT_least}.\footnote{Assumption \ref{hyp:strong_exogeneity} is not sufficient to ensure that DIDs comparing strongly and weakly treated units identify a causal effect: to obtain that result one needs to assume that the effect of receiving a low dose is the same in the two groups \citep{fricke2017identification}.}

\medskip
Under Assumption \ref{hyp:strong_exogeneity}, one can show that
\begin{equation}\label{eq:HAD_TWFE_dcDH2020_as}
\beta_{fe}=E\left(\frac{(D_2-E(D_2))D_2}{E\left((D_2-E(D_2))D_2\right)}E(\text{TE}_2|D_2)\right).
\end{equation}
Equation \eqref{eq:HAD_TWFE_dcDH2020_as} is an asymptotic version of the decomposition in Proposition S1 of \cite{dcDH2020}. It says that $\beta_{fe}$ is a weighted sum of the conditional average slopes (CAS) $E(\text{TE}_2|D_2=d)$, across all values that $D_2$ can take, where  $E(\text{TE}_2|D_2=d)$ receives a weight proportional to $(d-E(D_2))d$. Therefore, under Assumption \ref{hyp:strong_exogeneity}, $\widehat{\beta}_{fe}$ is not consistent for WAS or $E[\TE_2]$ and it does not even converge to a convex combination of CAS: some weights in \eqref{eq:HAD_TWFE_dcDH2020_as}  are necessarily negative, since $P(0<D_2<E(D_2))>0$. We then consider the following assumption:

\begin{hyp}
	(Homogeneous and linear effect) There is a real number $\delta$ such that \\ $E(\text{TE}_2|D_2)=\delta.$ \label{hyp:HAD_conditionunbiasedTWFE}
\end{hyp}

If Assumptions \ref{hyp:strong_exogeneity} and \ref{hyp:HAD_conditionunbiasedTWFE} hold, $\beta_{fe}=\delta=\WAS$. The following conditions are sufficient for Assumption \ref{hyp:HAD_conditionunbiasedTWFE} to hold:
\begin{align}
&E\left(\frac{Y_2(d)-Y_2(0)}{d}\middle|D_2=d\right)=E\left(\frac{Y_2(d)-Y_2(0)}{d}\right),\label{eq:SC1_forTWFEvalidity}\\
&E\left(\frac{Y_2(d)-Y_2(0)}{d}\right)=\delta,\label{eq:SC2_forTWFEvalidity}
\end{align}
for almost all $d\in\Supp(D_2)$.
\eqref{eq:SC1_forTWFEvalidity} is an homogeneous effect assumption, which assumes that the effect of moving treatment from 0 to $d$ is the same for units with different treatment doses.
\eqref{eq:SC2_forTWFEvalidity} is a linear effect assumption. Thus, slightly abusively, we refer to  Assumption \ref{hyp:HAD_conditionunbiasedTWFE} as an homogeneous and linear effect assumption.

\subsection{Assumptions \ref{hyp:strong_exogeneity} and \ref{hyp:HAD_conditionunbiasedTWFE} have a testable implication}\label{subsec:TWFEtest}

Let $\beta_0=E(\Delta Y)-\beta_{fe}E(D_2)$ denote the intercept of the TWFE regression.
\begin{thm}\label{thm:HAD_testability_robustnessTWFE}
\begin{enumerate}
\item In Design \ref{des:HAD}, if Assumptions \ref{hyp:strong_exogeneity} and \ref{hyp:HAD_conditionunbiasedTWFE} hold, then $E(\Delta Y|D_2)=\beta_0+\beta_{fe}D_2$.
\item In Design \ref{des:HAD_stayersorquasistayers}', if Assumptions \ref{hyp:bounded_slopes} and \ref{hyp:strong_exogeneity}  hold, $E(\Delta Y|D_2)=\beta_0+\beta_{fe}D_2$ implies that Assumption \ref{hyp:HAD_conditionunbiasedTWFE} holds and $\beta_{fe}=\delta$.
\end{enumerate}
\end{thm}

Based on Theorem \ref{thm:HAD_testability_robustnessTWFE}, we make the following recommendations.

\paragraph{If a linearity test of $E(\Delta Y|D_2)$ rejects, do not use $\widehat{\beta}_{fe}$.}
Point 1 of
Theorem \ref{thm:HAD_testability_robustnessTWFE} shows that if Assumptions \ref{hyp:strong_exogeneity} and \ref{hyp:HAD_conditionunbiasedTWFE} hold, $E(\Delta Y|D_2)$ is linear. By contraposition, if $E(\Delta Y|D_2)$ is not linear, then Assumptions \ref{hyp:strong_exogeneity} and \ref{hyp:HAD_conditionunbiasedTWFE} cannot hold. Then, we recommend not using $\widehat{\beta}_{fe}$. A caveat of that recommendation is that Assumptions \ref{hyp:strong_exogeneity} and \ref{hyp:HAD_conditionunbiasedTWFE} are not necessary to have that $\beta_{fe}$ is consistent for a causal effect. For instance, $\beta_{fe}=E(\TE_2)$ if
\begin{equation}\label{eq:nocorr_dose_deltaY(0)}
\Cov(\Delta Y(0),D_2)=0,
\end{equation}
\begin{equation}\label{eq:CNS_consistencyTWFE}
\Cov\left(\frac{(D_2-E(D_2))D_2}{E\left((D_2-E(D_2))D_2\right)},E(\text{TE}_2|D_2)\right)=0,
\end{equation}
two conditions that are respectively weaker than Assumption \ref{hyp:strong_exogeneity} and \ref{hyp:HAD_conditionunbiasedTWFE}. Yet, we are not aware of a test of \eqref{eq:nocorr_dose_deltaY(0)} and \eqref{eq:CNS_consistencyTWFE}.\footnote{If the data contains another pre-period $t=0$ where units are all untreated, as in period $t=1$, one can run a pre-trend test of \eqref{eq:nocorr_dose_deltaY(0)}, for instance by regressing $Y_{1} - Y_{0}$ on $D_{2}$, because $Y_{1} - Y_{0}=Y_{1}(0) - Y_{0}(0)$ is an outcome evolution without treatment. On the other hand, we are not aware of a test or placebo test that researchers can use to assess the plausibility of \eqref{eq:CNS_consistencyTWFE}.} This reduces the appeal of invoking \eqref{eq:nocorr_dose_deltaY(0)} and \eqref{eq:CNS_consistencyTWFE} to justify $\beta_{fe}$: tests or placebo tests are important to substantiate the credibility of one's findings.

\paragraph{If a linearity test of $E(\Delta Y|D_2)$ and a pre-trend test of Assumption \ref{hyp:strong_exogeneity} do not reject, one may use $\widehat{\beta}_{fe}$.}
%Assumption \ref{hyp:bounded_slopes} is a weak, technical condition.
If the data contains another pre-period $t=0$ where units are all untreated, as in period $t=1$, one can run a pre-trend test of Assumption \ref{hyp:strong_exogeneity}, by assessing if $Y_{1} - Y_{0}$ is mean-independent of $D_{2}$, because $Y_{1} - Y_{0}=Y_{1}(0) - Y_{0}(0)$ is an outcome evolution without treatment. Then, in designs with a QUG, Point 2 of Theorem \ref{thm:HAD_testability_robustnessTWFE} shows that under Assumptions \ref{hyp:bounded_slopes} and \ref{hyp:strong_exogeneity}, there is an ``if and only if'' relationship between the homogeneous-and-linear-effect condition in Assumption \ref{hyp:HAD_conditionunbiasedTWFE} and the linearity of $E(\Delta Y|D_2)$. This leads us to propose the following estimation rule. First: i) test that $E(\Delta Y|D_2)$ is linear; ii) conduct a pre-trend test of Assumption \ref{hyp:strong_exogeneity}; iii) test the null that there is a QUG. If none of these tests are rejected and the tests have decent power, one may use $\widehat{\beta}_{fe}$. Below, we discuss the consequences of these pre-tests on inference. If the tests in i) and ii) do not reject but that in iii) does, one could still consider using $\widehat{\beta}_{fe}$, acknowledging that one has only been able to test an implication of Assumption \ref{hyp:HAD_conditionunbiasedTWFE}.

\subsection{Testing that $E(\Delta Y|D_2)$ is linear, and pre-trend tests of Assumption \ref{hyp:strong_exogeneity}.}\label{subsec:stute}

%\paragraph{Nonparametric and tuning-parameter-free tests when $D_2$ takes a finite number of values.}
%Assume that $D_2$ takes $K$ values. If $K=2$, we necessarily have $E(\Delta Y|D_2)=\beta_0+\beta_{fe}D_2$, so there is no room for testability. If $K>2$, to test that $E(\Delta Y|D_2)$ is linear one can just regress $\Delta Y_g$ on a constant, $D_{g,2}$, $D_{g,2}^2$, ... $D_{g,2}^{K-1}$, and test that the coefficients on $D_{g,2}^2$, ..., $D_{g,2}^{K-1}$ are all zero. One can proceed similarly to test that $Y_{1} - Y_{0}$ is mean-independent of $D_{2}$, except that one tests that the coefficients on $D_{g,2}$, ..., $D_{g,2}^{K-1}$ are all zero, and now there is room for testability even when $K=2$.

\paragraph{A nonparametric and tuning-parameter-free test of linearity.}
We rely on \cite{stute1997nonparametric} to test the linearity of $E(\Delta Y|D_2)$. Under the null hypothesis that $E(\Delta Y|D_2)$ is linear, then $(\widehat{\eps}_{\text{lin},g})_{g=1,...,G}$, the residuals of the linear regression of $\Delta Y_g$ on $D_{g,2}$, should not be correlated with any function of $D_2$. Then, consider the following marked process:
$$c_G(d):=G^{-1/2} \sum_{g=1}^G \ind{D_{g,2}\le d} \widehat{\eps}_{\text{lin},g}.$$
\cite{stute1997nonparametric} shows that under the null hypothesis, $c_G$
converges to a Gaussian process. On the other hand, under the alternative, $c_G(d)$ tends to infinity for some $d\in\Supp(D_2)$. Then, one can consider Kolmogorov-Smirnov or Cram\'er-von Mises test statistics based on $c_G(d)$. We use a Cram\'er-von Mises test statistic:
\begin{equation}\label{eq:stat_Stutetest}
S := \frac{1}{G} \sum_{g=1}^G c^2_G(D_{g,2}).
\end{equation}
The limiting distribution of $S$ under the null is complicated, but \cite{stute1998bootstrap} show that one can approximate it using the wild bootstrap (see Appendix \ref{app:Stute} for details). \cite{stute1997nonparametric} and \cite{stute1998bootstrap} show that the test has asymptotically correct size, is consistent under any fixed alternative, and has non-trivial power against local alternatives converging towards the null at the $G^{-1/2}$ rate.

\paragraph{Testing the linearity assumptions underlying event-study effects.}
With more than two periods, if treatment occurs at $t=F\ge2$ and $D_t=D_F$ for all $t\ge F$, one can compute several event-study estimators, by regressing $Y_t - Y_{F-1}$ on $D_F$. As in Theorem \ref{thm:HAD_testability_robustnessTWFE}, one can show that the parallel-trends and homogeneous-and-linear-effect assumptions underlying these estimators have a testable implication: they imply that $E[Y_t-Y_{F-1}|D_F]=\beta_{0,t}+\beta_{fe,t}D_F$. To test this implication, one can use a test statistic similar to that in \eqref{eq:stat_Stutetest}, replacing the residuals from a  regression of $Y_2-Y_1$ on a constant and $D_2$ by the residuals from a regression of $Y_t - Y_{F-1}$ on a constant and $D_F$ in the definition of $S$.

\paragraph{Pre-trend tests of Assumption \ref{hyp:strong_exogeneity}: tests of mean independence.}
With more than two periods, and if treatment occurs at $t=F\ge3$, one can conduct a pre-trend test of Assumption \ref{hyp:strong_exogeneity}, by assessing if $Y_t - Y_{F-1}$ is mean-independent of $D_F$, for any $t< F-1$.
In the definition of the test statistic in \eqref{eq:stat_Stutetest}, one just replaces the residuals from a  regression of $Y_2-Y_1$ on a constant and $D_2$ by the residuals from a regression of $Y_t - Y_{F-1}$ on a constant.

\paragraph{Joint Stute tests.}
With more than two periods, one may be able to compute more than one linearity test of $E[Y_t-Y_{F-1}|D_F]$ for $t\ge F$, and more than one mean-independence test of $E[Y_t-Y_{F-1}|D_F]$ for $t< F-1$. In Appendix \ref{app:Stute}, we propose a new joint test that $E[Y_t-Y_{F-1}|D_F]$ is linear for all $t\ge F$, and a new joint test that $E[Y_t-Y_{F-1}|D_F]$ is constant for all $t<F-1$. The \st{stute\_test} Stata and R \citep[see][]{stute_test} commands compute all these tests.

\paragraph{Pre-testing and inference.} We recommend pre-testing that $E[\Delta Y|D_2]$ is linear and that $E[Y_1-Y_0|D_2]=E[Y_1-Y_0]$ before using the TWFE estimator. Theorem \ref{thm:HAD_testability_robustnessTWFE} and the results in  \cite{chaisemartin2024} imply that if, indeed, $E[\Delta Y|D_2]$ is linear and $E[Y_1-Y_0|D_2]=E[Y_1-Y_0]$, and Assumptions \ref{hyp:bounded_slopes} and \ref{hyp:strong_exogeneity} hold, inference conditional on accepting the two tests is still valid.

\section{Application: effect of bonus depreciation on employment}\label{sec:appli}

\paragraph{Research question and data.}
%Bonus depreciation has become a prominent tool for the US government to stimulate investment since the 2000s.
In the 2002 Job Creation and Worker Assistance Act, the US government introduced a bonus depreciation regulation,
%was introduced, which later got increased to 50, and even up to 100 percent for a short episode in 2011. This policy allows
whereby firms can deduct around 30 percent of the purchase price of a new investment from their taxable income in the year the investment is made, yielding a decrease in the present value of investment costs. That decrease is larger for longer-lived assets: as those assets depreciate more slowly, an immediate tax deduction is more valuable in present value as otherwise deductions would have been realized further in the future. Based on this insight, \citet{garrett_tax_2020} generate a county-level exposure measure $D_g$, equal to county $g$'s labor force share that works in industries relying on long-lived assets.
Our sample is the data used to produce Panel A of Figure 2 in the paper, a panel with employment at the county-industry-year level from 1997 to 2012. As exposure is at the county-year level, to fit in the setting considered in this paper we aggregate employment at the county-year level. $Y_{g,t}$ is the growth of employment in county $g$ from 2001 to $t$, and $D_{g,t}=D_g 1\{t\geq 2002\}$.

\paragraph{There are untreated and quasi-untreated counties in this application.}
12 counties are such that $D_g=0$, so the null hypothesis $\underline{d}=0$ is verified. However, the number of untreated counties is too low to use them only as the control group. One could also drop those 12 counties for fear that they differ too much from the overall population. We keep them, but dropping them yields results similar to those below. After dropping untreated counties, our test of the null hypothesis $\underline{d}=0$ is not rejected: $D_{2,(1)}=0.044$, $D_{2,(2)}=0.069$, and $T=1.77$, thus yielding a p-value of 0.361. Therefore, there are both untreated and quasi-untreated counties in this application.

\begin{figure}[t]
\caption{Effects of Bonus Depreciation on Employment \label{fig:fig_garrett}}
\begin{center}
\includegraphics[width = 0.8\textwidth]{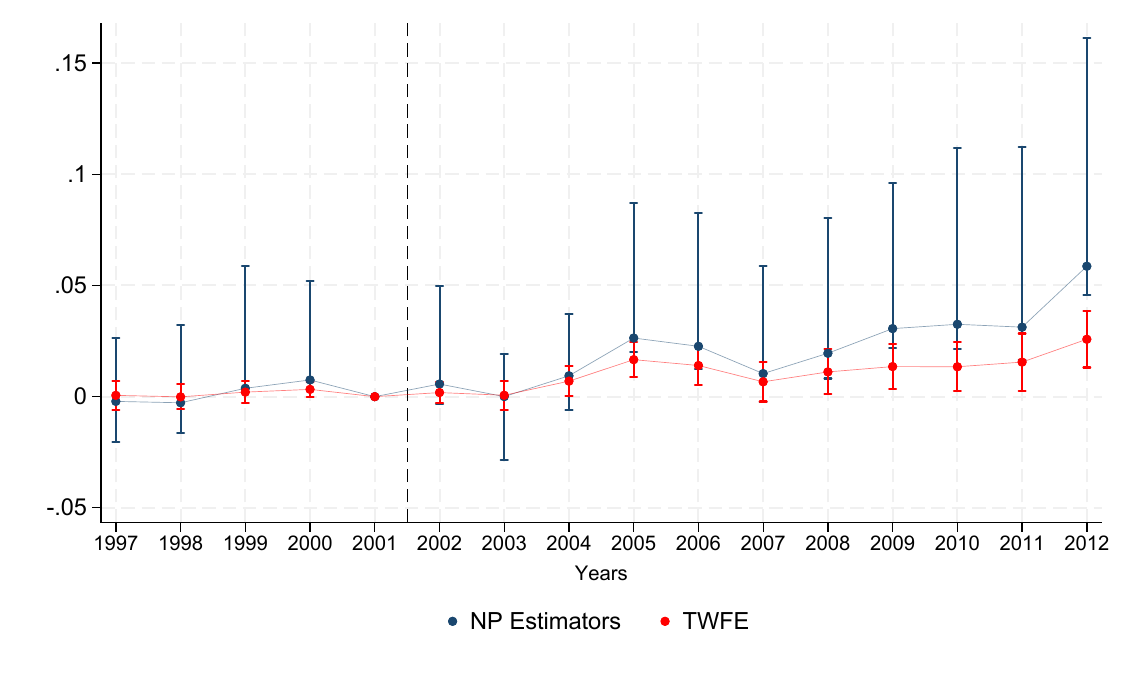}
     \begin{minipage}\textwidth
    \footnotesize{\textit{Notes:} Figure \ref{fig:fig_garrett} shows nonparametric and TWFE event-study estimators of the effect of bonus depreciation on employment, computed following Equation \eqref{eq:beta_QS}, using a 1997 to 2012 panel of 2,954 US counties constructed from the data of \cite{garrett_tax_2020}. The figure also shows pre-trend estimators, as well as estimators' 95\% level confidence intervals, which in the nonparametric case are computed with a bias-correction following Equation \eqref{eq:CI_QS}.}
    \end{minipage}
\end{center}
\end{figure}

\paragraph{Nonparametric event-study estimates indicate a positive effect of the treatment on employment.}
Figure \ref{fig:fig_garrett} below shows nonparametric event-study and pre-trend estimators, and their bias-corrected confidence intervals (BCCIs). Those estimators and BCCIs are computed as described in Sections \ref{sub:estimation_WAS} and \ref{sub:extensions}. For instance, at $t=2002$ (resp. $t=2003$, $t=2004$...), the event-study estimator is computed as $\widehat{\beta}^{\text{np}}_{\widehat{h}_G^*}$, defining the outcome variable as $Y_{g,2002}-Y_{g,2001}$ (resp. $Y_{g,2003}-Y_{g,2001}$, $Y_{g,2004}-Y_{g,2001}$...). Pre-trend estimators are small. They are marginally significant and positive in 2000 and in 1999, thus suggesting that prior to the reform, employment was increasing less on average than in the least-treated counties. Assuming that such differential trends would have continued after the reform, event-study estimators are downward biased. Starting in 2004, event-study estimators are positive and either significant or marginally insignificant, thus suggesting a positive effect of the reform on employment. Owing to the bias correction, confidence intervals are not centered around the point estimates.

\paragraph{TWFE estimates are twice smaller than the nonparametric estimates, and the Stute test is rejected.}
TWFE estimates, computed by regressing  $Y_{g,t}-Y_{g,2001}$ on $D_g$, are also shown on Figure \ref{fig:fig_garrett}. They are twice smaller than the nonparametric estimates.\footnote{Our TWFE estimates are close to those on Panel A of Figure 2 in \citet{garrett_tax_2020}.} A joint Stute test that $E(Y_{g,t}-Y_{g,2001}|D_g)$ is linear in $D_g$ for all $t\in \{2002,...,2012\}$ is strongly rejected (p-value<0.0001). A pre-trend test of Assumption \ref{hyp:strong_exogeneity}, assessing if $E(Y_{g,t}-Y_{g,2001}|D_g)$ is mean-independent of $D_g$ for all $t\in \{1997,...,2000\}$, is also strongly rejected (p-value<0.0001). Thus, the assumptions underlying the TWFE estimators are rejected in this application. At the same time, these estimators are also much more precise than the nonparametric estimators, so their bias might be offset by their precision. Overall, the nonparametric estimators and TWFE estimators complement each other, and it is reassuring that they share the same sign.

\section{Conclusion}

We study difference-in-differences designs in which all units are initially untreated and subsequently receive heterogeneous, strictly positive treatment doses. We show that in designs with a quasi-untreated group, a weighted average treatment effect per unit of treatment is identified under parallel trends by comparing the outcome evolution of the full population to that of quasi-untreated units. We develop a nonparametric estimator, bias-corrected inference, pre-trend and event-study estimators, and a test of whether a quasi-untreated group exists.

\medskip
The nonparametric approach is robust to treatment-effect heterogeneity but converges more slowly than the usual TWFE estimator. We therefore also study when TWFE can be justified. We show that parallel trends together with an homogeneous and linear effect assumption imply a testable linearity restriction on the conditional mean of outcome changes. This leads to a practical specification-testing strategy for deciding when TWFE may be credible.

\medskip
In our application to bonus depreciation, the nonparametric estimates imply substantially larger effects than the corresponding TWFE estimates. More generally, our results suggest that quasi-untreated units can provide a useful source of identification in settings where no unit remains fully untreated, while specification tests can help assess when simpler TWFE methods are adequate.

\newpage
\putbib

\end{bibunit}

\newpage
\appendix

\begin{bibunit}

\newpage
\setcounter{page}{1}
\begin{center}
    {\huge Supplemental appendix, not for publication}
\end{center}

\section{Additional material on the non-parametric estimator}\label{sec:add_np}

\subsection{Regularity conditions} % (fold)
\label{sub:regularity_conditions}

Let $m(d) := E[\Delta Y|D_2 = d]$ and $\sigma^2(d) := V(\Delta Y|D_2 = d)$.
Optimal bandwidth $\hat{h}^*_{G}$ is computed using the procedure described in Section 3.3 of \cite{calonico2018effect}.
Theorem 2(c), Corollary 4(b) and Corollary 5(b) therein guarantee that the confidence interval in \eqref{eq:CI_QS} has the desired asymptotic coverage for WAS, provided that the following assumptions are satisfied.
We refer to \cite{calonico2018effect} and the Supplement therein for a discussion of each assumption.

\begin{hyp}[Regularity conditions]
In what follows, values and derivatives at zero are interpreted as right-hand limits.
\begin{enumerate}
    \item \textbf{Data generating process.} $D_2$ has an absolutely continuous distribution with density $f_{D_2}$. There exists $\delta>0$ such that $E[|\Delta Y|^{8+\delta}|D_2] < \infty$, uniformly over $\Supp(D_2)$.
    \item \textbf{Local polynomial regressions.} Estimation is carried out via a local polynomial regression of degree $p \geq 1$ and kernel $K(.)$. Similarly, bias correction is carried out via a local polynomial regression of degree $q = p + 1$ and kernel $L(.)$. Kernels $K(.)$ and $L(.)$ are positive, bounded, even functions, with compact support.
    \item \textbf{Local regularity.} There exists some constant $d_0 > 0$ such that in $[0,d_0)$
    \begin{itemize}
        \item[a.] $f_{D_2}$ and $\sigma^2$ are continuous and bounded away from zero;
        \item[b.] $m$ is $S$ times continuously differentiable, with $S > q + 2$;
        \item[c.] the $S$-th derivative of $m$ is H\"older continuous, i.e. there exists some finite constant $C \geq 0$  and $\varsigma > 0$ such that    $|m^{(S)}(d)-m^{(S)}(d')|\leq C|d-d'|^{\varsigma}$ for all $d,d'\in[0,d_0)$.
    \end{itemize}
    \item \textbf{Bandwidths.} Let $\eta_{bc}$ be the bias of the bias corrected estimator,\footnote{The explicit form of $\eta_{bc}$ can be obtained by using the Edgeworth expansion reported in Section S.II.1.2 (p.47) of the Supplement to \cite{calonico2018effect}, and replacing $(Y,X,x,n)$ with $(\Delta Y, D_2, 0, G)$.} $\{h_G\}$ the sequence of bandwidths for estimation, and $\{b_G\}$ for bias correction. As $G \to \infty$, it holds that
    \begin{itemize}
        \item[a.] $h_G \to 0$, $b_G \to 0$, $h_G/b_G \to \bar{\rho} <\infty$ and $\bar{\rho} > 0$
        \item[b.] $G h_G / \log(G h_G) \to \infty$, $\eta_{bc} \log(Gh_G) \to 0$
    \end{itemize}
    \item  \textbf{Cram\'er condition.}\footnote{Point 5 of Assumption \ref{hyp:regul}, when all other points of Assumption \ref{hyp:regul} are satisfied, implies the Cram\'er condition from \cite{calonico2018effect}. See Assumption S.II.3 of the Supplement to \cite{calonico2018effect} for a formal statement of the Cram\'er condition, and Remark S.II.3 for a discussion of why Point 5 of Assumption \ref{hyp:regul} is sufficient.} For $u \in \R$, let $r_p(u) = (1, u, u^2, ..., u^p)'$ and $w(u) = \frac{1}{2}(u + 1)(u + 2)$.
    Let $\psi:[-1,1] \to \R^\kappa$, with $\kappa = 1 + w(p) + w(q)$, be defined as
    $$
    \psi(u) =
    \begin{pmatrix}
    1, \text{vech}\left(K(u)r_p(u)r_p(u)'\right)', \text{vech}\left(L(u)r_q(u)r_q(u)'\right)'
    \end{pmatrix}'
    $$
    for $u \in [-1,1]$, and let $\psi_j:[-1,1]\to \R$ be the projection of $\psi$  on the $j$-th column, for $j = 1, ..., \kappa$. The set of functions $\{\psi_j\}_{j =1}^\kappa$ is linearly independent on the space of continuous functions defined on $[-1,1]$.
\end{enumerate}
    \label{hyp:regul}
\end{hyp}

By default, \texttt{nprobust}, hence \texttt{did\_had}, use $p=1$, $q=2$ and Epanechnikov kernels.

% subsection regularity_conditions_underlying (end)

\subsection{Simulations}\label{sub:sims}

Table \ref{tab:sim_table} below shows the results from a simulation study, where we assess the coverage rate of the bias-corrected confidence interval (BCCI) based on \eqref{eq:CI_QS} and computed by \st{did\_had}.

We consider two DGPs. In DGP 1, $D_2$ follows a uniform distribution on $[0,1]$, $\Delta Y(0)$ follows a standard normal independent of $D_2$, and $\Delta Y_{2}(D_2) = D_2 + D_2^2 + \Delta Y(0)$, thus implying that $\text{WAS}=5/3$. Assumptions \ref{hyp:PT_least}, \ref{hyp:bounded_slopes}, and \ref{hyp:regul} hold in this DGP, so the BCCI should have close to nominal coverage when the sample size is large enough. When $G=100$, $\hat{\beta}^{np}_{\hat{h}^*_G}$ is slightly upward biased, and the 95\% level BCCI slightly undercovers, with a coverage rate of 89\%. Its coverage rate increases to 93\% when $G=500$, and to 95\% when $G=2500$.

\medskip
DGP 2 is similar to DGP 1, except that now $D_2$ is drawn from a Beta distribution $B(2,2)$, thus implying that WAS$=8/5$. The rationale for DGP 2 is to test coverage in a setting where Assumption \ref{hyp:regul} fails because $f_{D_2}(0) = 0$.
With a coverage rate of 90\%, the 95\% level BCCI slightly undercovers when $G=100$, but its coverage almost reaches 95\% when $G=2500$. Thus our BCCI can still have close to nominal coverage when $f_{D_2}(0) = 0$.

\begin{table}[H]
    \caption{Simulation Results}
    \label{tab:sim_table}
%\scriptsize
%\resizebox{\linewidth}{!}{
\centering
\aboverulesep=0ex\belowrulesep=0ex
\iffalse\begin{tabular}{lcccc}
\toprule 
& \multicolumn{2}{c}{DGP 1} &  \multicolumn{2}{c}{DGP 2} \\ 
\cmidrule(lr){2-3}\cmidrule(lr){4-5}
& (1) & (2) & (1) & (2) \\ 
\midrule 
G = 100&    1.6850&    0.8875&    1.6526&    0.8985 \\ 
G = 500&    1.6962&    0.9265&    1.6320&    0.8960 \\ 
G = 2500&    1.6832&    0.9510&    1.6313&    0.9445 \\ 
\bottomrule 
\end{tabular}\fi
\begin{tabular}{lcccc}
\toprule 
& \multicolumn{2}{c}{DGP 1} &  \multicolumn{2}{c}{DGP 2} \\ 
\cmidrule(lr){2-3}\cmidrule(lr){4-5}
& (1) & (2) & (1) & (2) \\ 
\midrule 
G = 100&    1.69&    0.89&    1.65&    0.90 \\ 
G = 500&    1.70&    0.93&    1.63&    0.90 \\ 
G = 2500&    1.68&    0.95&    1.63&    0.94 \\ 
\bottomrule 
\end{tabular}

%}
\\
     \begin{minipage}\textwidth
    \footnotesize{\textit{Notes:} Column (1) shows $\hat{\beta}^{np}_{\hat{h}^*_G}$, computed following Equation \eqref{eq:beta_QS}. Column shows (2) the coverage of the bias-corrected 95\%-level CI, computed following Equation \eqref{eq:CI_QS}. The two DGPs are described in the text. 2,000 simulations were conducted per DGP/sample size.}
    \end{minipage}
\end{table}

\section{Further details on the Stute (1997) test}\label{app:Stute}

\paragraph{Wild bootstrap.} % (fold)
\label{par:wild_bootstrap}

Consider testing the null that $E(\Delta Y|D_2)$ is linear in $D_2$. We obtain a given bootstrap sample (indexed by $b$)  by first drawing $G$ i.i.d. realizations $\{\xi^b_g\}_{g = 1}^G$ from the two-point distribution in \cite{mammen1993bootstrap} and then letting $\Delta Y^*_{g,b} = \Delta Y_g + (\xi^b_g - 1) \widehat{\eps}_{\text{lin},g}$. The bootstrap sample is then $(D_{g,2},\Delta Y^*_{g,b})_{g=1,...,G}$. Let $S^*_b$ denote the $S$ statistic obtained on this sample. Repeating this over $B$ bootstrap samples, we finally obtain the the p-value of the test by $B^{-1}\sum_{b=1}^B 1\{S < S^*_b\}$.

% paragraph wild_bootstrap (end)

\paragraph{Joint test with $T>2$ periods.} % (fold)
\label{par:joint_test_with_t_2_periods}

As mentioned in the text, we can define a similar statistic $S^t$ for testing the linearity of $E[Y_t-Y_{F-1}|D_F]$ when $t\ge F$. Then, we also develop a joint test that linearity holds for all $t\ge F$. The test statistic is $S^{\geq F} = \sum_{t \geq F} S^t$. We compute its p-value as above, using bootstrap samples $(D_{g,F-1},...,D_{g,T},\Delta Y^*_{g,F,b},..., \Delta Y^*_{g,T,b})_{g=1,...,G}$ obtained as follows. We first draw $\{\xi^b_g\}_{g = 1}^G$ as above. Then, we let $\Delta Y^*_{g,t,b} = Y_{g,t}-Y_{g,F-1} + (\xi^b_g - 1) \widehat{\eps}_{\text{lin},g,t}$, with $\widehat{\eps}_{\text{lin},g,t}$ the residual from the linear regression of $Y_t-Y_{F-1}$ on $D_F$. Following \cite{cameron2008bootstrap}, we consider the same $\xi^b_g$ for $t=F,...,T$ to preserve the serial correlation in the true residuals. With bootstrap sample $b$, we then compute $(S^{*,t}_b)_{t=F,...,T}$ and in turn $S^{*,\geq F}_b= \sum_{t \geq F} S^{*,t}_b$. Finally, we consider a similar test of the hypothesis that $E[Y_t-Y_{F-1}|D_F]$ is constant for all $t< F-1$.

% paragraph joint_test_with_t_2_periods (end)

\paragraph{Simulations on the joint test.} % (fold)
\label{par:sim_stute}

We consider $T=6$, $F=2$ and the following DGP
\begin{equation}\label{eq:dgp_stute}
    Y_{g,t}(D_{g,t})=\alpha_g + \gamma_t + \theta_t \cdot D_{g,t} + \eps_{g,t},
\end{equation}
in which $D_{g,t}=D_g\ind{t\ge F}$ and $D_g$ is uniformly distributed on $[0,1]$. Group fixed effects satisfy $\alpha_g|D_g\sim\mathcal{N}(1+D_g,1)$, time fixed effects satisfy $\gamma_t = 0.1 \times t$ and $(\eps_{g,t})_{t\ge 1}$ follows a normal, stationary AR(1) process with autocorrelation $0.5$ and variance $4/3$. The treatment effect $\theta_t$ is homogeneous across groups but varies with time, $\theta_t = 0.25 \times (t-1)$. We  test that linearity holds for all $t\ge F$. Since this is the case in this DGP, the test should reject approximately at the nominal level $\alpha$. Table \ref{tab:sim_table_stute} confirms it for $\alpha\in\{0.01, 0.05, 0.1\}$ and $G\in\{100, 500, 1,000\}$.

\begin{table}[H]
    \caption{Simulation Results for the joint Stute Test}
    \label{tab:sim_table_stute}
%\scriptsize
%\resizebox{\linewidth}{!}{
\centering
\aboverulesep=0ex\belowrulesep=0ex
\begin{tabular}{lccc}
\toprule 
& $\alpha = 0.01$ & $\alpha = 0.05$ & $\alpha = 0.1$ \\
\midrule 
G = 100&    0.011&    0.053&    0.103 \\ 
G = 500&    0.009&    0.055&    0.108 \\ 
G = 1000&   0.009&    0.049&    0.104 \\ 
\bottomrule 
\end{tabular}
\\
     \begin{minipage}\textwidth
    \footnotesize{\textit{Notes:} Each column reports the rejection rates of the joint Stute test at the nominal level $\alpha$. The DGP is described in the text. We use 1,000 simulated samples and $B=500$ bootstrap samples for each sample.}
    \end{minipage}
\end{table}

\section{Proofs}

\subsection{Theorem \ref{thm:HAD_WaldDID}}

By Assumption \ref{hyp:bounded_slopes},
\[
|Y_2(D_2)-Y_2(0)|\le M D_2,
\]
so conditional on \(D_2\le d\),
\[
E\!\left[|Y_2(D_2)-Y_2(0)|\mid D_2\le d\right]\le Md\to0.
\]
Since
$$E[\Delta Y|D_2\le d] = E[\Delta Y(0)|D_2\le d] + E[Y_2(D_2)-Y_2(0)| D_2\le d],$$
we get, by Assumption \ref{hyp:PT_least}, that $\lim_{d\downarrow 0} E\left[\Delta Y|D_2\le d\right]=E[\Delta Y(0)]$. Moreover,
\begin{align*}\label{eq:HAD_thm_step1}
E[\Delta Y]-\lim_{d\downarrow 0} E\left[\Delta Y|D_2\le d\right]=&E[Y_2(D_2)-Y_1(0)]-E[Y_2(0)-Y_1(0)]\nonumber\\
=&E[Y_2(D_2)-Y_2(0)].
\end{align*}
The result follows from the previous display and \eqref{eq:HAD_WaldDID}.

\subsection{Theorem \ref{thm:test_inf_support}} % (fold)
\label{sub:proof_of_theorem_ref_thm_test_inf_support}

\subsubsection*{1. Asymptotic size control.} % (fold)
\label{ssub:point_1}

Fix a sequence $(x_G)_{G\ge 1}$ with $x_G>0$ and $x_G\to 0$; without loss of generality, we assume hereafter that $x_G\le m\overline{d}$ for all $G$. Fix also  a sequence $(F_G)_{G\ge 1}$, with $F_G\in \mathcal{F}^{0,\overline{d}}_{m,K}$ for all $G\ge 1$. Consider any subsequence $(F_{G'})_{G'\ge 1}$ of $(F_G)_{G\ge 1}$. Since $F_G'$ is Lipschitz with parameter $K$ on $[0,\overline{d}]$, the subsequence $(F'_{G'})_{G'\ge 1}$ is  equicontinuous and $\sup_{x\in[0,\overline{d}]} F'_G(x)\le 1/\overline{d}+K\overline{d}$.  Then, by Arzel\`a–Ascoli theorem, it admits a further subsequence that converges uniformly; let $H_1(\cdot)$ denote its limit. By the mean value theorem (and since $F_{G''}^{-1}(x_{G''})\le \overline{d}$),
$$F_{G''}(F_{G''}^{-1}(x_{G''}))= F'_{G''}(d_{G''}) F_{G''}^{-1}(x_{G''}),$$
for some $d_{G''}\in [0, F_{G''}^{-1}(x_{G''})]$. Hence, $F'_{G''}(d_{G''})\ne 0$ and
$$\frac{F_{G''}^{-1}(x_{G''})}{x_{G''}} = \frac{1}{F'_{G''}(d_{G''})}.$$
Because $F^{-1}_G(u) \le u/m$ for all $u\in[0,m\overline{d}]$, $F_{G''}^{-1}(x_{G''})\to 0$, which implies that $d_{G''}\to 0$. Then, by uniform convergence of $F'_{G''}$ and $H_1(0)\ge m$ (in view of $F_G'(0)\ge m$), we obtain
$$\frac{F_{G''}^{-1}(x_{G''})}{x_{G''}} \to \frac{1}{H_1(0)}.$$
For $U$ uniformly distributed over $[0,1]$, $F_G^{-1}(U) \eqd D_2$. As a result, by the representation of spacings \citep[see, e.g.,][p.721]{ShorackWellner},
\begin{equation}
\left(D_{2,(1)}, D_{2,(2)}-D_{2,(1)}\right) \eqd \left(F_G^{-1}\left(\frac{E_1}{\sum_{i=1}^{G+1} E_i}\right), F_G^{-1}\left(\frac{E_1+E_2}{\sum_{i=1}^{G+1} E_i}\right) - F_G^{-1}\left(\frac{E_1}{\sum_{i=1}^{G+1} E_i}\right)\right),	
	\label{eq:repres}
\end{equation}
where $(E_1,...,E_{G+1})$ are i.i.d. and follow an Exponential(1) distribution. Also, by the law of large numbers,
$$\frac{E_1}{\sum_{i=1}^{G+1} E_i} \convP 0, \quad \frac{E_1+E_2}{\sum_{i=1}^{G+1} E_i} \convP 0.$$
Hence, by the extended continuous mapping theorem \citep[see, e.g.,][Theorem 18.11]{vanderVaart2000},
\begin{align*}
& G'' \bigg(F_{G''}^{-1}\bigg(\frac{E_1}{\sum_{i=1}^{G''+1} E_i}\bigg), \;  F_{G''}^{-1}\bigg(\frac{E_1+E_2}{\sum_{i=1}^{G''+1} E_i}\bigg) - F_{G''}^{-1}\bigg(\frac{E_1}{\sum_{i=1}^{G''+1} E_i}\bigg)\bigg) \notag \\
= & \; \bigg(\frac{E_1}{\frac{1}{G''}\sum_{i=1}^{G''+1} E_i} \times \frac{F_{G''}^{-1}\bigg(\frac{E_1}{\sum_{i=1}^{G''+1} E_i}\bigg)}{E_1/\sum_{i=1}^{G''+1} E_i}, \frac{E_1+E_2}{\frac{1}{G''}\sum_{i=1}^{G''+1} E_i} \times \frac{F_{G''}^{-1}\bigg(\frac{E_1+E_2}{\sum_{i=1}^{G''+1} E_i}\bigg)}{(E_1+E_2)/\sum_{i=1}^{G''+1} E_i} \notag \\
& \; - \frac{E_1}{\frac{1}{G''}\sum_{i=1}^{G''+1} E_i} \times \frac{F_{G''}^{-1}\bigg(\frac{E_1}{\sum_{i=1}^{G''+1} E_i}\bigg)}{E_1/\sum_{i=1}^{G''+1} E_i}\bigg) \notag \\
\convP & \frac{1}{H_1(0)} \times (E_1, E_2). %\label{eq:conv_for_as_size}
\end{align*}
Combined with \eqref{eq:repres} and, again, the continuous mappping theorem, we obtain
\begin{equation}
\frac{D_{2,(1)}}{D_{2,(2)}-D_{2,(1)}} \convL \frac{E_1}{E_2},
	\label{eq:conv_for_as_size}
\end{equation}
where the convergence should be understood along the previous subsequence. Then,
$$P_{F_{G''}}(W_\alpha)  =  \; P_{F_{G''}}\left(\frac{D_{2,(1)}}{D_{2,(2)}-D_{2,(1)}}>1/\alpha - 1\right) \to  \alpha.$$
Now, let $u_G:=P_{F_G}(W_\alpha)$. The previous display proves that $u_{G''}\to \alpha$. By Urysohn subsequence principle, this implies that $u_G \to \alpha$. Since the sequence $(F_G)_{G\ge 1}$ was arbitrary, $\lim\sup_{G\to\infty} \sup_{F\in \mathcal{F}^{0, \overline{d}}_{m,K}} P_F(W_\alpha) = \alpha$.

\subsubsection*{2. Uniform consistency} % (fold)
\label{ssub:consistency}

If $F\in\mathcal{F}^{\underline{d},\overline{d}}_{m,K}$ for some $\underline{d}>0$, we have
\begin{equation}
T = \frac{\underline{d} + \tilde{D}_{2,(1)}}{\tilde{D}_{2,(2)} - \tilde{D}_{2,(1)}},	
	\label{eq:repr_T_power}
\end{equation}
where the cdf of $\tilde{D}_{2}:=D_2-\underline{d}$ belongs to $\mathcal{F}^{0,\overline{d}-\underline{d}}_{m,K}$. We reason as above by considering sequences $F_G $ in $\mathcal{F}^{0,\overline{d}-\underline{d}}_{m,K}$ and appropriate subsequences for which $F^{(k)}_G$ converges uniformly.
By the same reasoning as above, $(\tilde{D}_{2,(1)}, \tilde{D}_{2,(2)})\convP 0$, along such subsequences. By \eqref{eq:repr_T_power}, this implies that, still along such subsequences, $T\convP \infty$. Point 2 follows by Urysohn subsequence principle again.

% subsubsection consistency (end)

\subsubsection*{3. Local power} % (fold)
\label{ssub:3_local_power_with_positive_density_at_0}

As above, consider a sequence $F_G$ of cdfs in $\mathcal{F}^{\underline{d}_G,\overline{d}}_{m,K}$. % and we prove that $u_G:=\min(0,P_{F_G}(W_\alpha)-\alpha) \to 0$.
For any subsequence, we consider a further subsequence ($G''$) such that (i) the  derivative of the cdf of $\tilde{D}_{2,k}$ converges and (ii) the sequence $(G''\underline{d}_{G''})$  converges; we denote by $\lambda\in [c,\infty]$ its limit. Then, using \eqref{eq:repr_T_power} but with $\underline{d}$ replaced by $\underline{d}_G$ and the same reasoning as to get \eqref{eq:conv_for_as_size}, we obtain, after some algebra and along the subsequence ($G''$),
$$T \convL T_\lambda:=\frac{\lambda H_1(0)+ E_1}{E_2},$$
with the understanding that $T_\infty=\infty$. In this latter case, the same reasoning as in Point 2 above shows that $P_{F_{G''}}(W_\alpha)\to 1$. Now, assume that $\lambda< \infty$. First, remark that for any $q, x>0$,
$$P\left(\frac{q+ E_1}{E_2}>x\right) = 1-\frac{x}{x+1}\exp(-q/x).$$
Then, using $H_1(0)\ge m$ and $\lambda\ge c$, we obtain
\begin{align*}
P(T_\lambda>1/\alpha-1)= & 1- (1-\alpha)\exp\left(-\frac{\alpha \lambda H_1(0)}{1-\alpha}\right)\\
\ge & 1- (1-\alpha)\exp\left(-\frac{\alpha c m}{1-\alpha}\right) =:\alpha^*,
\end{align*}
with $\alpha^*>\alpha$. By Urysohn subsequence principle, this implies that
$$\liminf_G P_{F_G}(W_\alpha) \ge \alpha^* >\alpha.$$
The result follows since the initial sequence $F_G$ was arbitrary in the class $\mathcal{F}^{\underline{d}_G,\overline{d}}_{m,K}$.

% subsubsection 3_local_power_with_positive_density_at_0 (end)

%\subsection{Theorem \ref{thm:HAD_testability_robustnessTWFE_QS}.}
%Remark that
%\begin{equation}\label{eq:proof_thm:HAD_testability_robustnessTWFE_QS1}
%\Delta Y = Y_2(\underline{d}) - Y_1(0) + (D_2 - \underline{d}) \text{TE}_{2, \underline{d}}.
%\end{equation}
%Then,  under Assumptions \ref{hyp:strong_exogeneity} and \ref{hyp:homog_min}, one has that:
%\begin{equation}\label{eq:proof_thm:HAD_testability_robustnessTWFE_QS2}
%E[Y_2(\underline{d}) - Y_1(0)|D_2]=E[Y_2(\underline{d}) - Y_1(0)],
%\end{equation}
%an equivalent of Assumption \ref{hyp:strong_exogeneity} but for outcome evolutions under the lowest period-two treatment. Then, the proof is similar to that of Theorem \ref{thm:HAD_testability_robustnessTWFE}, replacing $D_2$ and TE$_2$ by $D_2-\underline{d}$ and $\text{TE}_{2, \underline{d}}$, respectively.

\subsection{Theorem \ref{thm:HAD_testability_robustnessTWFE}}
In Design \ref{des:HAD},
\begin{align}\label{eq:HAD_thm_step1}
E(\Delta Y|D_2)=&E(\Delta Y(0)|D_2)+D_2 E(\text{TE}_2|D_2)\nonumber\\
=&\mu_0+D_2 E(\text{TE}_2|D_2),
    \end{align}
where the second equality follows from Assumption \ref{hyp:strong_exogeneity}.
Point 1 of Theorem \ref{thm:HAD_testability_robustnessTWFE} directly follows from plugging Assumption \ref{hyp:HAD_conditionunbiasedTWFE} into \eqref{eq:HAD_thm_step1} and from the fact that if $E(U|V)=a_0+a_1V$ then $E(U|V)$ is equal to the linear regression of $U$ on $(1,V)$.

\medskip
Then, assume that we are in Design  \ref{des:HAD_stayersorquasistayers}' and
\begin{equation}\label{eq:HAD_thm_step2}
E(\Delta Y|D_2)=\beta_0+\beta_{fe} D_2.
\end{equation}
\eqref{eq:HAD_thm_step1} and \eqref{eq:HAD_thm_step2} imply
that
$$\beta_0 - \mu_0 = E(Y_2(D_2)-Y_2(0)|D_2) - \beta_{fe}D_2.$$
Hence, by the law of iterated expectations, for all $d$
$$\beta_0 - \mu_0 = E(Y_2(D_2)-Y_2(0)|D_2\le d) - \beta_{fe}E[D_2|D_2\le d].$$
We already showed that under Assumption \ref{hyp:bounded_slopes}, $\lim_{d\downarrow 0} E(Y_2(D_2)-Y_2(0)|D_2\le d) = 0$. Since we also have $\lim_{d\downarrow 0}E[D_2|D_2\le d]=0$, we obtain $\beta_0=\mu_0$.
Then, equating \eqref{eq:HAD_thm_step1} and \eqref{eq:HAD_thm_step2} implies that
\begin{align*}
  D_2 E(\text{TE}_2|D_2)=& \beta_{fe}D_2,
\end{align*}
and dividing by $D_2>0$ yields $E(\text{TE}_2|D_2)=\beta_{fe}$.
This implies that $E(\text{TE}_2|D_2)=E(\text{TE}_2)= \beta_{fe}$, which proves Point 2.

\putbib
\end{bibunit}

\end{document}